\documentclass[prl,twocolumn,showpacs,superscriptaddress,amsfonts]{revtex4}

\usepackage{psfig}

\begin{document}

\title{
Dynamic localization and Coulomb blockade in quantum dots under AC 
pumping}
\author{D.~M.~Basko}
\email{basko@ictp.trieste.it}
\affiliation{The Abdus Salam International Centre for Theoretical Physics,
Strada Costiera 11, 34100 Trieste, Italy}
\author{V.~E.~Kravtsov}
\affiliation{The Abdus Salam International Centre for Theoretical Physics,
Strada Costiera 11, 34100 Trieste, Italy}
\affiliation{Landau Institute for Theoretical Physics,
2 Kosygina Street, 117940 Moscow, Russia}

\date{\today}
\begin{abstract}
We study conductance through a quantum dot under Coulomb blockade
conditions in the presence of an external periodic perturbation.
The stationary state is determined by the balance between the
heating of the dot electrons by the perturbation and cooling by
electron exchange with the cold contacts. We show that Coulomb
blockade peak can have a peculiar shape if heating is affected by
dynamic localization, which can be an experimental signature of
this effect.
\end{abstract}


\maketitle

{\it Introduction.}---
Experimental observation of dynamic localization~(DL) in trapped
ultracold atoms in the field of a modulated laser standing
wave~\cite{Raizen} provided a solid ground for the preceding
extensive theoretical studies of the kicked quantum
rotor~\cite{Izrailev,Haake}.
In a recent publication~\cite{us} we have shown that an analogous
suppression of the energy absorption is possible for a solid-state
system -- a chaotic quantum dot under an ac excitation, e.~g. like
those used in experiments of Ref.~\cite{Marcus}, which makes the
question about the possibility of observation of DL in a quantum
dot highly relevant. If one wishes to detect this effect by
transport measurements, the Coulomb blockade
regime~\cite{MarcusRev,Aleinerrev} is the most suitable, since it
is in this regime that the transport is sensitive to the internal
state of the dot. For an open dot, when electron-electron
interaction can be neglected, the conductance is insensitive to
the electron energy distribution in the
dot~\cite{Vavilov,Kanzieper}.

Under non-equilibrium conditions the (effective) electronic
temperature~$T$ of the dot is determined by the balance between
heating by the ac perturbation and cooling due to various
mechanisms. At sufficiently low~$T$ cooling is dominated by
simple electronic exchange between the dot and the contacts (the
latter are assumed to be maintained at a constant low temperature
determined by the cryostat), while the energy exchange with the
phonon subsystem is negligible~\cite{phonons}.
In this case, as the gate voltage is detuned from the Coulomb
blockade peak, the cooling rate decreases, leading to an increase
of~$T$.
The dynamic localization manifests itself in suppression of
heating, which becomes $T$-dependent. This makes the shape of the
Coulomb blockade peak sensitive to dynamic localization.  

{\it Heating.}---
First, consider the standard picture of heating by an ac
perturbation.
Let the single-electron mean level spacing~$\delta$ in the dot
be small enough. Then, if an external time-dependent periodic
perturbation with the frequency~$\omega$ is applied, the total
electronic energy~$E$ in the dot (counted from that of the ground
state) grows linearly with time as described by the Fermi Golden
Rule: $E(t)=\Gamma\omega^2t/\delta\equiv{W}_0t$. The
probability per unit time~$\Gamma$ of each single-electron
transition measures the ac power~\cite{perturbation}. 
The criterion of validity of
the Fermi Golden Rule is $\delta\ll\Gamma$, and $\Gamma\ll\omega$
is also assumed ($\hbar=1$). 

This classical picture is valid only if each act of photon
absorption by an electron is independent of the previous ones;
however, for a discrete (though dense) energy spectrum 
this turns out not to be the case. After many transitions the
absorption rate decreases due to accumulation of the quantum
interference correction~\cite{Izrailev,Haake,us}, so that after
a time $t_*\sim\Gamma/\delta^2$ the absorption is completely
suppressed. This effect is known as the dynamic localization in
energy space; the effective electronic temperature
$T_*\sim\Gamma\omega/\delta$ (the characteristic spread of the
electron distribution function) reached by the time~$t_*$, plays
the role of the localization length. 
Note that DL has nothing to do with the saturation of absorption
by a pumped two-level system, as in our case the spectrum is
unbounded.

The considerations of Ref.~\cite{us} were based on random matrix
theory description of the single-particle properties of the dot,
valid provided that all energy scales in the problem are small
compared to the Thouless energy~$E_{Th}$
(defined by the order of magnitude as the inverse of
the time required for an electron to travel across the dot and
thus to randomize its motion due to scattering off the dot
boundaries).
For the dot to be in the Coulomb blockade regime, the effective
temperature should be also smaller than the dot Coulomb charging
energy~$E_c$. In the following, the hierarchy of scales
$\delta\ll\Gamma\ll\omega\ll{T}_*\ll{E}_{Th},E_c$ is assumed
(for a typical 2D GaAs dot $\delta\sim{1}$~$\mu$eV,
$E_{Th}\sim{100}$~$\mu$eV~$\sim{1}$~K~\cite{Marcus}; the
stronger the inequality $\delta\ll{E}_{Th}$ is satisfied, the
better).


Being an interference effect, DL requires perfect quantum
coherence.
Electron interaction and/or connection to leads causes electron
{\it dephasing}. In the presence of {\em weak} dephasing with
the rate~$\gamma_{\phi}\ll{1/t_*}$ there is a residual absorption
with the rate determined by~$\gamma_{\phi}$:
\begin{equation}\label{WinDL=}
W_{\rm{in}}\sim{W}_0\gamma_{\phi}t_*=
T_*^2\,\gamma_{\phi}/\delta\:.
\end{equation}
If the dephasing is too strong, $\gamma_{\phi}\gtrsim{1}/t_*$,
the dynamic localization is destroyed and $W_{\rm{in}}\sim{W}_0$.

The expression~(\ref{WinDL=}) was justified in Ref.~\cite{Basko}
for the dephasing due to electron-electron collisions provided 
that the effective electronic temperature $T>T_{*}$.
The main condition of its applicability is that dephasing should
be a sequence of distinct phase-destroying events with average
rate ~$\gamma_{\phi}$, rather than phase diffusion, so that the 
dephasing rate roughly coincides with the quasiparticle
relaxation rate: $\gamma_{\phi}\sim\gamma_{qp}$.
This is certainly correct for the case of electron escape to
the contacts, since in this case the electron is effectively
replaced by another one with an absolutely random phase.
This is also true for electron-phonon collisions and
electron-electron collisions in a quantum dot, since the typical
energy transfer during a collision is of the order of the
(effective) electronic temperature in the dot, which is large:
$T>T_{*}\gg{1/t_*},\gamma_{qp}$  ($T_*\gg{1}/t_*$ due to
$\Gamma,\omega\gg\delta$). 
Note that this condition is the reason why no exponentially small
factor like $e^{-T_{*}/T}$ or  $e^{-t_{*}^{-1}/T}$ arises in 
Eq.~(\ref{WinDL=}) in contrast to hopping conductivity in
disordered systems at low temperatures.

Once the condition $\gamma_{\phi}\sim\gamma_{qp}$ is verified,
the following consideration can be applied. As the collisions
are rare ($\gamma_{qp}t_*\ll{1}$), the electrons spend most of
the time in the states localized in the energy space, having
definite phase relationships. When at some moment the phase of
some electron is destroyed, its wave packet starts spreading
along the energy axis. It localizes again after the time
$\sim{t}_*$, in the meantime spreading by $\sim{T}_*$.
Thus, the ac driven dynamics following the collision leads to
a change of the total electronic energy of $\sim{T}_*$~per
collision. The sign of this change is, however, arbitrary,
because a periodic perturbation can equally cause transitions
up and down the spectrum. Only the presence of the filled Fermi
sea below (i.~e., an energy gradient of the electronic
distribution function) makes absorption the preferred direction,
which means that if the electronic temperature $T\gg{T}_*$, the
energy absorbed per collision is on the average $\sim{T}_*^2/T$
rather than~$T_*$. The effective number of electrons that can
participate in a collision is $\sim{T}/\delta$ (due to the
degenerate Fermi statistics).
During the time interval $\sim{1}/\gamma_{qp}$ each electron
participates in one collision, so the total number of
collisions per unit time is $\sim(T/\delta)\gamma_{qp}$.
This gives the energy absorption rate
$W_{\rm{in}}\sim({T}_*^2/T)(T/\delta)\gamma_{qp}$, which is
exactly Eq.~(\ref{WinDL=}).

\begin{figure}
\psfig{figure=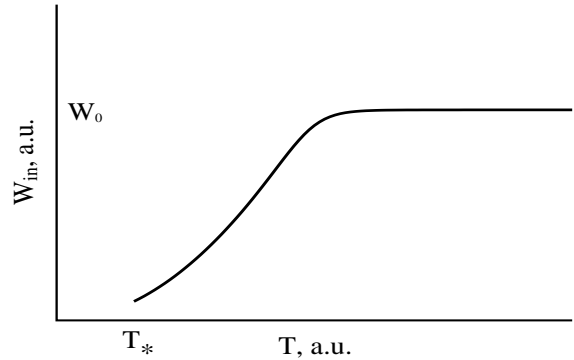,width=9cm,height=6cm}
\caption{\label{Win:}
A schematic view of the dependence of the absorption rate
$W_{\rm{in}}$ on the effective electronic temperature in the
dot: in the dynamic localization regime the absorption is
due to dephasing, so it is temperature-dependent; when the
temperature becomes high enough, the dephasing destroys the
DL. $T<T_*$ cannot be realized in the regime of strong DL.}
\end{figure}

The same can be seen from an alternative argument.
After each collision the electron spends the time $\sim{t}_*$
absorbing the energy from the microwave field, then it stops to
absorb (the DL occurs) and waits for the next event (provided
that $t_*\ll{1}/\gamma_{\phi}$). Thus the absorption rate of
the whole system is given by the simple weighted average:
$W_{\rm{in}}\sim{W}_0\gamma_{\phi}t_*$, which
is again Eq.~(\ref{WinDL=}).


As~$\gamma_{\phi}$, generally speaking, depends on the electronic
temperature, so does the absorption rate in the DL regime
(Fig.~\ref{Win:}).
The quasiparticle relaxation rate due to electron collisions in
a quantum dot was calculated by Sivan, Imry and Aronov~\cite{SIA}:
\begin{equation}\label{SIA=}
\gamma_{e-e}(T)\sim\delta\,(T/E_{Th})^2.
\end{equation}
The derivation of this
expression implies the effective continuity of the many-particle
spectrum, which imposes a condition
$T_*\gg\sqrt{E_{Th}\delta/\ln(E_{Th}/\delta)}$~\cite{AGKL}.
Obviously, for the dynamic localization to have any chance to
develop, the condition $\gamma_{e-e}(T_*)t_*\ll{1}$ should also be
satisfied.

{\it Sequential tunneling.}---
We characterize the coupling of the dot to two contacts by
electron escape rates $\gamma_1,\gamma_2$
($\gamma_1+\gamma_2\equiv\gamma\ll\delta$). Let $U$~be the energy
cost of adding an electron to the dot, proportional to the gate 
voltage (at the exact degeneracy point $U=0$, corresponding to
the very top of the Coulomb blockade peak).
When $U$~is not far from the degeneracy point, the main
contribution to the conductance comes from the leading
order of the perturbation theory in the dot-contact coupling.
For characteristic temperatures $T\gg\delta$ one can describe
the system by the rate equations of Kulik and
Shekhter~\cite{Shekhter}.

It is quite straightforward to consider these equations for
a general (non-equilibrium) electron energy distribution
function in the dot. As an estimate for the electronic
distribution function we use the Fermi-Dirac form with some
temperature~$T$. We also assume~$T$ to be much higher than the
temperature of the contacts~$T_{0}$ ($T_{0}$ can be made as
low as $\sim{10}$~mK~\cite{Kouwenhoven2003}), which is true
if the pumping power is high enough. Then a simple calculation
gives the following expressions for the dot conductance~$G$
(in units  $e^{2}/2\pi\hbar$), ``renormalized'' electron
escape rate~$\gamma_{esc}$, and the cooling
rate~$W_{\rm{out}}$ [we denote $x\equiv{U}/(2T)$,
$G_0\equiv\gamma_1\gamma_2/(\gamma\delta)$]:
\begin{eqnarray}
\label{GVG0=}
G(U)&=&G_0
\left[1-\frac{x\tanh{x}}{\ln(2\cosh{x})}\right],\\
\label{gescCBP=}
\gamma_{esc}(U)&=&
\frac{\gamma}{2}\left[1-\frac{|x|}{\ln(2\cosh{x})}\right],\\
\label{WoutCBP=}
W_{\rm{out}}(U)&=&\frac{\gamma}{\delta}\,T^2
\left[\frac{\pi^2}{12}-x^2+
\frac{2x \int\limits_0^x{y}\tanh{y}\,dy}{\ln(2\cosh{x})}
\right].
\end{eqnarray}

{\it Stationary state.}---
For each given~$U$ the temperature~$T$ of the stationary state
is found from the energy balance equation
$W_{\rm{out}}=W_{\rm{in}}$, where $W_{\rm{out}}$~is given
by Eq.~(\ref{WoutCBP=}), $W_{\rm{in}}$ -- by Eq.~(\ref{WinDL=}),
the dephasing rate -- by Eqs.~(\ref{gescCBP=}),~(\ref{SIA=}).
Substituting it into Eq.~(\ref{GVG0=}), one obtains the shape
of the Coulomb blockade peak.

\begin{figure}
\psfig{figure=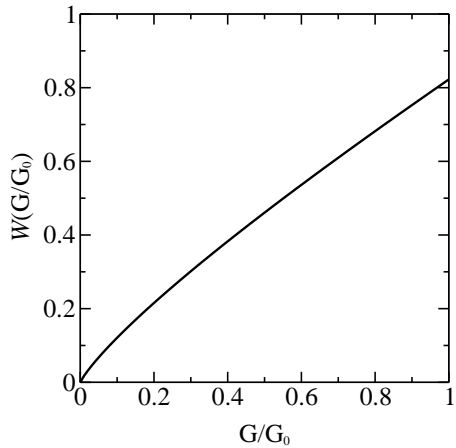,width=7cm,height=7cm}
\caption{\label{calW:}
The function $\mathcal{W}(G/G_0)$ defined in Eq.~(\ref{calW=}).}
\end{figure}

Suppose for a moment that dephasing is dominated by
electron-electron collisions, while cooling is dominated
by the escape to the contacts~\cite{phonons}.
One can notice the following property of Eqs.~(\ref{GVG0=})
and (\ref{WoutCBP=}): $G/G_0$ and
$W_{\rm{out}}(U)/[(\gamma/\delta)T^2]$ are functions of
$x\equiv{U}/(2T)$ only. This allows us to write a relation
\begin{equation}\label{calW=}
\frac{W_{\rm{out}}}{T^2}=(\gamma/\delta)\,\mathcal{W}(G/G_0)\,.
\end{equation}
The function $\mathcal{W}(G/G_0)$ is plotted in
Fig.~\ref{calW:} and with logarithmic precision we have 
$(\gamma/\delta)\,\mathcal{W}(G/G_0)\sim G$. Thus the physical 
meaning of Eq.~(\ref{calW=}) is similar to the Wiedemann-Franz
law. 

The energy balance condition takes the form
\begin{equation}
W_{\rm{in}}\sim{T}_*^2\,(\gamma_{\phi}(T)/\delta)=
W_{\rm{out}}=(\gamma/\delta)\,T^2\,\mathcal{W}(G/G_0)\,,
\end{equation}
or 
$G\sim 
(\gamma/\delta)\mathcal{W}(G/G_0)=(\gamma_{\phi}(T)/\delta)\,
(T_*/T)^2$. 
Remarkably, for $\gamma_{\phi}(T)$ given by Eq.(\ref{SIA=})
$U$~and~$T$ drop out and the solution of this equation
for~$G$ is independent of~$U$, leading to a flat plateau
$G\sim(T_*/E_{Th})^2$ on the Coulomb blockade curve $G(U)$.  
Note that the 
the plateau conductance must be smaller than the peak conductance
$G_{0}\sim \gamma/\delta$. 
Therefore, the solution exists only if
\begin{equation}\label{conditionopen=}
\gamma/\delta\gtrsim
(T_*/E_{Th})^2.
\end{equation}

Now consider the top of the peak, $U=0$. Including the
dephasing due to both escape and electron-electron collisions,
we can write the energy balance condition as
\begin{equation}\label{balanceU0=}
(\gamma/\delta)\,T^2\sim(\gamma/\delta)\,T_*^2+
(T_*/E_{Th})^2T^2.
\end{equation}
Here the left-hand side represents the cooling rate in the peak,
the two terms on the right-hand side come from the dephasing
due to escape and electron collisions, respectively. Due to the 
condition~(\ref{conditionopen=}) the second term is smaller than
the left-hand side, so the only way to satisfy the equation is to
have $T(U=0)\sim{T}_*$. Thus, for the dynamic localization to be
possible the dephasing in the very peak of the Coulomb blockade
{\em must} be dominated by escape.
One can extract the temperature of the
stationary state at $U=0$ measuring the curvature of the peak,
and study its dependence on control parameters: intensity~$\Gamma$
and coupling to the contacts~$\gamma$.

\begin{figure}
\psfig{figure=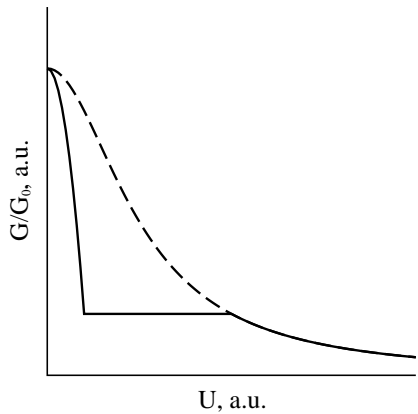,width=7cm,height=7cm}
\caption{\label{CBscheme:}
A sketch of the Coulomb blockade peak shape in the DL regime
(solid line): at small~$U<U_{\rm{min}}$ the dephasing is
dominated by the electron escape (peak), at larger~$U$ --
by electron-electron collisions (plateau), and finally, at
$U>U_{\rm{max}}$ the cooling is insufficient, the dynamic
localization is destroyed, and the dot is in the Ohmic regime.
The Ohmic curve is also shown for reference by the dashed line.}
\end{figure}

As $U$~is detuned from the peak,
the dot becomes effectively more closed. Thus, the crossover from
the peak to the plateau occurs when the two mechanisms are
equally efficient. With logarithmic precision this happens at
\begin{equation}\label{beginplateau=}
T\sim{T}_*\,\quad U\sim{U}_{\rm{min}}\sim{T}_*\:.
\end{equation}
The plateau ends when the temperature of the dot becomes so large
that the dynamic localization is destroyed by dephasing and
$W_{\rm{in}}\sim{W}_0$. Obviously, this happens
when the plateau $G\sim (T_*/E_{Th})^2$
hits the curve $G_{D}(U)\sim (W_{0}/U^{2})\,\ln(G_{0}U^{2}/W_{0})$ 
determined from 
the condition of classical (Ohmic) absorption
$W_0=W_{\rm{out}}$. With logarithmic precision this happens at
\begin{equation}\label{endplateau=}
U\sim{U}_{\rm{max}}\sim{E}_{Th}\sqrt{\delta/\Gamma}\:.
\end{equation}
The resulting shape of the Coulomb blockade peak is drawn
schematically in Fig.~\ref{CBscheme:}.

\begin{figure}
\psfig{figure=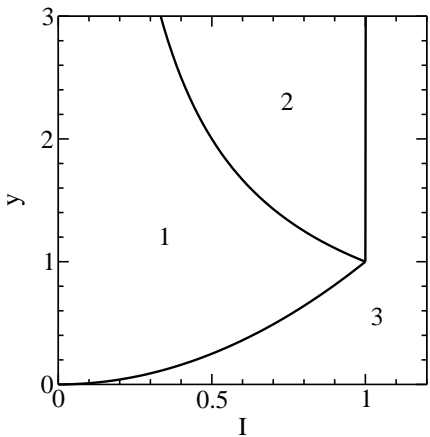,width=7cm,height=7cm}
\caption{\label{phasediag:}
A schematic view of the ``phase diagram'' in terms of the
dimensionless intensity and escape rate ($I-y$ plane), without
taking into account cooling and dephasing due to phonons.
The top of the Coulomb blockade peak corresponds to DL only in
the region~1; the flat plateau in the tails exists both in 
regions $1$~and~$2$; in the region~3 DL is absent.}
\end{figure}

It is convenient to introduce two dimensionless parameters,
corresponding to two experimentally controllable parameters
$\Gamma$~(ac power)and~$\gamma$ (tunneling into leads):
\begin{equation}
I\equiv\frac{\Gamma}{\delta}
\left(\frac{\omega}{E_{Th}}\right)^{2/3},\quad
y\equiv\frac{\gamma}{\delta}
\left(\frac{\omega}{E_{Th}}\right)^{-2/3}.
\end{equation}
The necessary condition for dynamic localization 
$\gamma_{e-e}(T_*)t_*\ll{1}$ becomes $I\ll{1}$,
the condition~(\ref{conditionopen=}) is $y\gg{I}^2$. The top
of the peak will correspond to DL regime if $\gamma{t}_*\ll{1}$
or $Iy\ll{1}$~\cite{plateau}. The resulting ``phase diagram''
is shown in Fig.~\ref{phasediag:}.

{\it Conclusions.}---
We have studied electronic conduction through a quantum dot in
the Coulomb blockade regime under an external periodic perturbation.
In contrast to the well-studied equilibrium case, the electronic
temperature of the ac~driven dot is determined by the balance
between heating by the perturbation and cooling due to electron
exchange with cold contacts. The cooling rate thus depends on the 
gate
voltage, and so does the dot temperature. As the gate voltage is
detuned away from the peak, the cooling rate decreases, and the
temperature increases.
In the strong dynamic localization regime the heating rate is
determined by dephasing, as the usual linear absorption is
blocked by quantum interference. 
The most peculiar situation is realized when the dephasing is
due to electron-electron collisions: in this case the Coulomb
blockade peak has a flat shoulder, where the conductance does
not depend on the gate voltage. Such a shape could be an
experimental signature of the dynamic localization effect.

We are grateful to Yu.~M.~Galperin, B.~N.~Narozhny, C.~M.~Marcus,
and B.~L.~Altshuler for helpful discussions.


\begin{thebibliography}{99}
\bibitem{perturbation}
The transition rate is proportional to the power of the microwave
and can be estimated as $\Gamma\sim(e\mathcal{E}L)^2/E_{Th}$,
where $\mathcal{E}$~is the amplitude of the electric field in
the dot, $L$~is the dot size, and $E_{Th}$ is the Thouless energy.
If the screening length $1/\kappa<L$, one should substitute
$\mathcal{E}L\rightarrow\mathcal{E}_{ext}/\kappa$.
Finally, instead of using microwave one can change the dot shape
modulating the gate voltage.
\bibitem{plateau}
Note that to observe of the plateau the condition
$\gamma\ll{1}/t_*$ {\em is not} necessary: even if at $U=0$
the DL is absent, as $U$~is increased, the dot becomes
effectively more closed, so the dephasing by escape becomes
less efficient.
\end{thebibliography}
\end{document}